**Walter Hehl**

# A General World Model with Poïesis:
# Popper's 'Three Worlds' updated with Software.

Walter Hehl, Thalwil, former IBM Research, Zürich, Switzerland.
Seestrasse 5, CH-8800 Thalwil, Switzerland.
Walter.H.Hehl@gmail.com
Phone: +41447201544

**Abstract**
With the famous 'Three Worlds' of Karl Popper as template, the paper rigorously introduces the concept of software to define the counterpart of the physical subworld. Digesting the scientific-technical view of biology and neurology on a high level, results in an updated 'Three Worlds' scheme consistent with an information technical view. Chance and mathematics complete the world model. Some simple examples illustrate the move from Popper's view of the world with physics, psyche and *World 3,* to a new extended model with physics, extended software (which we call Poïesis), and 'Geist' (the notion which embodies spirit, mind and soul).

## 1. An Introduction to Popper's 'Three Worlds'

In his famous paper and Tanner lecture 'Three Worlds' [1], and in several other books, the Austrian-British philosopher Karl Popper outlines his view of the universe by distinguishing three sub-universes. He denominates these sub-universes as *Worlds* [1].

He considers *World 1* as the physical world. It consists of the stars and stones, of radiation and all other forms of energy. He also includes plants and animals in this class of soulless objects, commenting that living vs. non-living could possibly be regarded as two subclasses of *World 1*. His view is obviously totally human-centric.

Secondly, he defines a human world, his *World 2*, of mental states and psychology. Here he lists 'objects' such as feelings of pain and pleasure, thoughts and decisions, perceptions and observations. He understands *World 2* as the realm of subjectivity, with psychological states as feelings and with thought processes as decisions.

Most important to Karl Popper is the proof of existence and the defense of what he calls *World 3*, which for him is the world of the products of the human mind. Popper's *World 3* includes 'objects' such as languages and stories and religious myths, scientific theories and mathematical constructions, songs and symphonies, paintings and sculpture. He explicitly excludes from the higher level *World 3* the engineering products of mind, such as 'airplanes and airports'. On the other hand, he considers computer programmes presumably as *World*

---
[1][1] We denote the Popper worlds with capital initial.

*3* because he regards them as extended 'plans of action', and therefore as pure human products.

Popper demonstrates this, among others, with Beethoven's Fifth Symphony as a piece of art: Notes and specific performances of this symphony are *World 1* representations of the symphony per se. He regards the creation of Beethoven itself as a *World 3* object, namely THE 'Fifth Symphony of Beethoven'.

As a mundane example, we classify the cooking of a dish such as 'Zürich Ragout' following Popper: The recipe itself belongs to *World 3,* as does the language of the instructions. Recipe and language are products of the human mind. On the other hand, the ingredients are physical objects. Below we will revisit these examples in the light of this paper.

We regard the Popper 'Ansatz' with *World 1* (objective) – *World 2* (subjective) – *World 3* ('geistig') as a valid anthropomorphic classification of the world. But over the past seventy years a major technology close to us humans has emerged, and starts to tackle human capabilities: the computer or better, as we will explain, large software systems. Software is about to close the classical gap between "matter" and "Geist" corresponding to the gap between science and technology on one side and the humanities on the other. Software has changed from a free give-away when buying computer hardware, to the stuff which drives a car and 'eats the world'. We can use the notion of software to reclassify the world, technology and nature, in a dynamic way.

## 2. What is Software?

Software is defined as the set of instructions for a computer (or for another software) to perform a specific task. The (digital) computer consists of a concerted triple set of hardware for execution, architecture for the appropriate organization, and software for the fulfilment of a specific task. (Digital) software demonstrates a couple of properties of fundamental philosophical value: Software incorporates three completely different phases of action with different philosophical meaning.

First, the definition of the given task by writing the 'specs'. This describes the purpose or sense of the software. This is the view literally 'from above'.

Second, the phase of generation and build-up of the set of instructions; the software development. Compared to the construction of hardware, software is much easier to create and change, and is therefore in practice (incl. biology and neurology, see below) the carrier of complexity. This means that all complex systems are intrinsically software systems, even a computer chip with semiconductor circuits is de facto software; physics contributes mainly some miniaturised boundary layers for electrons and holes. The complexity of a system is measured by the size of the smallest software programme solving the problem (George Chaitin, 2012). This definition avoids redundancy to be counted.

Third, software operates on data and executes commands; it 'runs'. In our context, this is equivalent to being alive after receiving a slight push to start the command chain to run like a row of upright dominoes toppling over. This is the spark of life. Life and death are elements of subworld 2 including near death, which is experienced during the breakdown of the running software.

Meaning emerges from data through software; without software, data are only physics. Software systems do not just describe states, but provide continuous services in computers, in factories and in the human brain. Such streaming software processes of input and output data are reactions. A modern technical example is the software or "consciousness" of a Self-Driving car.

We generalise the notion of software further and define: *'Software is a set of instructions to a corresponding hardware which break momentarily the physical status of a system.'*
The momentary status can be the status of a flip-flop or switch, the position of an atom or a molecular protein base pair, the magnetic field of a particle, the integrity of a surface, the electric current through a membrane, a clean kitchen with the ingredients for a recipe waiting to be used, the upright position of dominos waiting to topple over, etc. Some of these elements can be used for computing in the mathematical sense, although rather unconventional.

Imagine a simple switch to turn on the light, just in position 'light off'. From the position of the switch, the hand of the operator appears out of nowhere and changes the position to 'on'. A more impressive example is a rocket flying in space with its motor off under the laws of gravitation; it is an object of subworld 1. The mission control centre sends a command to ignite, and it transitions to subworld 2.

In a running computer chip this interference happens millions or even billions of times a second with each machine instruction. We call this breaking in a 'downwards causation' with a notion coined by the philosopher and psychologist Donald T. Campbell in 1974. Observe that apart from these interventions, everything follows the causal laws of nature. A computer is therefore a device to perform complex sequences of 'downward causations'. In a spectrum ranging from sharp and digital operations in normal computers to fuzzy and sparkling biological computing, the software for digital computers is a well understood limiting case that enables the study of the laws of complex software systems.

The light switch is the simplest possible example just for didactics. Unfortunately, the didactic examples of (freshman) programming are philosophically disastrous, because they give the student the impression that software is a quasi-mechanical clockwork. This is not true, most notably by what we call the 'Scaling Law': Large software systems become more like organisms through the internal diversity and plurality, interactions and real time dependencies. Only those who have never had an error on their computer could disagree!

Another formulation of the false argument 'a computer is just a clockwork', is 'a computer cannot create really new', where 'new' can mean 'surprising the creator' or 'not thought of', or even 'not conceivable' by the creator when writing the programme. This feature is closely connected to the philosophical notion of emergence, the appearance of 'new' from parts not having this property at all. We consider that software can paint pictures and compose music (since 60 years), recognize faces, develop new shapes of aircraft wings, and can learn all by itself: The rules might be given by the programmer, or just the rules to make new rules. Software can change itself by learning from experience or by chance[2] (or both). The best example here is natural evolution and its laws for software development to 'higher and higher' life forms, usually with higher software complexity. Evolution is a broad software development with balanced variability and stability of code in interaction with a system of other evolutionary subsystems. This sounds complicated, but evolutionary software development is widely used to solve technical problems.

---

[2] The importance of chance will be elaborated below.

Identifying evolution as a grand software system development raises human questions, such as 'who made the specifications?', 'who developed the code' and 'who runs the system'? This evokes new philosophical, religious and human versions of panpsychism, pantheism, panentheism respectively panendeism and pandeism. Miracles, for example, would be interventions in the runtime environment, either in a weak form within the laws inhibited as chance ('Zufall'), or as 'strong' miracles violating the laws of nature or changing them on the spot. The famous 'Intelligent Design' claims interventions in the development phase.

Software often comes in a hierarchy of levels, implying upper levels closer to the visible surface of objects, such as, for example, the well-known 'Apps', and lower levels only accessible to the professional and closer to the 'bottom'; the respective hardware. These levels are called firmware or microcode. This multilayer concept as a guideline for the construction of large systems, is important for many applications.

In order to open up the understanding of software, we emphasise that software does not imply 'being digital' per se: Software can operate with blurred information, and even fuzzy 'hardware' and diffuse architecture. On a digital computer, this fuzziness can of course be simulated, although probably awkward, i.e. with many sharp instructions to perform one fuzzy one.

**3. A New Dualism with Extended Software (Poïesis)**
In our *World* concept, we model the physical subworld and the software subworld as neighbouring twin towers. The form factor 'tower' emphasises the erection of the physical world on an axiomatic base as the standard model of particle physics via atoms and molecules up to stars and galaxies. The software tower reflects the growth and buildup of software from small programmes to large systems respectively from simple organisms 'up to us humans'. In the sense of our extended software notion, we include in the software pillar notably three main areas:
(Information) technology, biological systems and the processes in the brain.

The software tower comprises all actions and processes that are 'made', i.e. deviate from the flow of non-living nature with its simple processes such as free motion in space, diffusion or crystallization and snow flake growth. We think that the term poïesis from the ancient Greek ποίησις 'to make', respectively the adjective poïetic, is an appropriate philosophical designation for the software subworld. In fact, the notion has been used by several philosophers from Plato to Martin Heidegger[3]. Here poïesis describes a well-defined quasi-technical process, which is well understood and even reasonably measurable in technical software engineering and through the architecture of digital computers.

---
[3] For Martin Heidegger, poïesis refers to a moment of change as the " coming-out of a butterfly from a cocoo".
 For us, poïesis implies the whole process of reproduction and growth of the animal.

Physical subworld with matter and energy, and poïetic subworld with software and processes, are stacked side-by-side but not completely independent from each other: The poïetic tower needs some physical objects as a base. Software operations at the base level, such as 'add two numbers' or 'store a word of information' or 'recognize a pattern', need some atoms or molecules, some electric charges or currents, magnetic fields, on/off or photon states. The minimum amount of energy required is the subject of physical-nature philosophical investigations, as started by Rolf Landauer (1961). In order to visualise the dependency of poïesis on matter and energy, we use an L-shaped graphical form for the physical subworld in figure 1; the software tower resides on the bar of the letter L.

The two towers are interacting and interactively developing with time. A large scale example from nature is the generation of oxygen by plants (poïetic world) and the atmosphere of the earth (physical world). A striking modern example is 3D-Printing: commands from the computer produce quite complex physical objects on a 3D-print facility. A more intricate interweaving of subworld 1 and 2, of body and soul, is known as embodiment.

In this so far dualistic model, our example 'cooking a Zürich ragout' is analysed as follows: The ingredients are subworld 1, the recipe is visibly software (subworld 2). In addition to the visible software, the cook activates internal software as subroutines, learned in his or her infancy, or at the culinary school; in computer language firmware or microcode.

Finally, we propose for the two worlds simple English (or Germanic) terms. Stimulated by the language of Poul Anderson in 'Uncleftish Beholding' (1989), we think it natural to call the *World 1* (physics and more) as 'Worldstuff', the *World 2* (extended software with genetics, brain processes and digital technical software) as 'Worldware'.

**4. The Updated *World 3* of the 'Geist'**
Important objects classified by Karl Popper as *World 3* can be explained as software objects, many of them undoubtedly because they have already been successfully replicated or handled by technical software systems. Languages and knowledge can be understood as large meshed networks of data and processes. Issues such as the limited quality of, for example, translations, are more practically caused by the sheer size of the task. Many pieces of art are proper software, both in generation as in reproduction, maybe with some random internals to induce creativity. This is even valid for Beethoven's Fifth Symphony which Popper used as an example. What else is the score of the symphony but a parallel programme on paper? The musicians add their firmware and microcode to handle the sophisticated degrees of freedom of their muscles and instruments. If the software-technical view on language is humiliating for the reader, just consider the translation of a washing machine manual as a soft technical task. This translation is no doubt a subworld 2 object, and translation is technically a software process.

But many of us have the feeling that there is more in the world than the two scientific-technical towers and their objects, namely 'objects' like 'genuine' art, deeply felt art, love and passion. As the questionable attributes indicate, we leave here the domain of substantial predicates, and we transcend from subworld 2 or claim the existence of respective transcendence. Love for example has certainly components from subworld 1 (such as hormones and smooth skin) and subworld2 (psychology, evolutionary roots, rational reasons), but what more is it? Most of us will insist there is a 'more'. We relate this 'more' to the German notion 'Geist', probably in the sense of the 'absolute Geist', as with Hegel. 'Geist' embodies the English terms 'spirit', 'mind' and 'soul', as well as the French 'esprit'[4] with meanings in different areas such as psychology (subworld 2), or even in chemistry as alcohol (subworld 1). Here we define 'Geist' as the answers to questions such as: 'What is really fine art?'. We get tangled up in questions such as 'Is my amateurish painting art? Is the much better painting produced by a computer art? Is this painting by Pablo Picasso art?', or 'What qualifies Hamlet to be a piece of world literature?', 'Why is Beethoven's Fifth Symphony so outstanding?', 'What is love more than the drive to reproduction and breeding?'

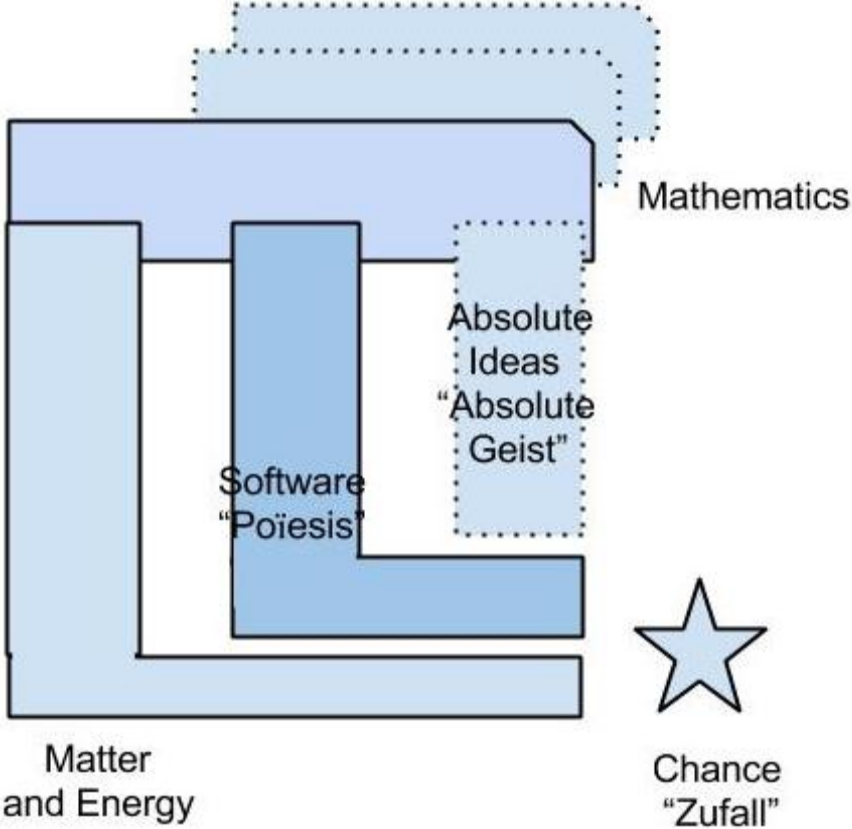

**Fig. 1   World Model with Three Worlds including Software for Poïesis,
with Chance (Zufall) on the outside, and with Mathematics.**

---

[4] These paraphrases for ‚Geist' are from the Hegel network *wiki.Hegel.net* drawn 2015.

For Beethoven's Fifth, we finally get a breakdown of the components:
Paper, acoustics (subworld 1), composition, harmony rules, handling of instruments (subworld 2) and potentially some ingenious and intangible 'Geist' (subworld 3) [5].

The 'Geist' of music respectively other varieties of 'absolute Geist' emerge (if existing) from products of the mind in subworld 2; in the case of the symphony, for example, from music theory, emotions and chance. We illustrate this in the figure by shaping the poïetic tower, also as for the physical L, by a capital L, with the 'Geist' residing above this bar.

For completeness, we try to identify the 'Geist' in technical software. From personal experience we compare this with the making of an invention. There are two acts of higher (probably 'geistiger') work required for a patent: first, the identification of a specific new problem, and second, the idea of the solution itself. In this sense, some 'Geist' is both in the specification, as well as in the idea of the invention.

## 6. Addenda: Chance and Mathematics

The world model has so far a fundamental flaw by suggesting a world only following laws, rules and programmes. Benoît Mandelbrot pointed to the established fact that the world is shaped by chance: Trees, the clouds, the waves on the sea, the position of stars in galaxies – they are all determined by laws of nature plus lots of random actions or 'Zufällen' within the frame of these laws. This German word 'Zufall' (plural 'Zufälle') is emotionally neutral and signifies some breaking and intruding action, with unknown causes, into certain situations. The causes might be hidden (cryptodeterministic) or not existing at all (quantum 'Zufall'). An individual tree is generated by the execution of some relatively small genetic software, plus a vast amount of 'Zufall' for the details, such as the shapes of leaves and the configuration of branches. This implies that the overwhelming part of data in the world has been inserted by 'Zufall'. Regular geometrical objects (such as large crystals) or simple analytics, are more the exception than the rule. Because software per se is also data, this also introduces random software in the world. With this we have three fundamental modes of interaction in the world:
       Causal, downwards causal and by accident or chance ('zufällig').
Therefore we add the 'Zufall' to our world model, and annex to the subworlds in figure 1 a large star symbol to acknowledge the importance of random processes and the penetration of 'Zufall' into the world. As a supplementary comment: this could be interpreted as interventions by some higher power from outside, but of course perfectly within the world and in the framework of the laws and rules of the subworlds.

We attach to the world model a mysterious fifth element: Mathematics. Technically, mathematics generates many worlds, but among these is the real world. Mathematics reflects even features in the atomic or cosmic world which are ungraspable for us humans (Walter Hehl, 2012). Albert Einstein in 1921 expressed his feeling regarding the unbelievable mapping of mathematics into nature (or vice versa) as 'How can it be that mathematics, being after all a product of human thought which is independent of experience, is so admirably appropriate to the objects of reality?'. For a physicist, it is indeed a miracle that the theory of mathematical groups from the 19th century defines the 230 different crystallographic types for minerals in

---
[5] Below we add chance (Zufall) as a possible means for creativity.

the world; or some functions defined (invented or detected) by the French mathematician Adrien-Marie Legendre in the 18th century, reflect the periodic system of chemical elements and the clouds of electrons around atoms and molecules. Mathematics is related to all three subworlds:

In physics, a large part of mathematics has even been developed relating to software. Mathematics can be interpreted as software itself, both as operations or proofs; and regarding the 'Geist', we could see mathematical principles (e.g. the prime numbers or even number theory in general) themselves as 'Geist'. This is compliant with Popper's view.
Finally, for taking into account 'Zufall' there is a dedicated field of mathematics - Statistics.

## 7. Final Conclusions

The world model as presented here mainly differs in three aspects from the *Worlds* of Karl Popper.

First, Popper classified the world mainly in non-human effects vs. human products. We divide (or better construct) the world on one hand in non-living natural courses in the physical subworld 1, and on the other hand in the made or constructed subworld 2, described as extended software or as poïesis. The main emphasis is therefore on processes either as physical (causal only) flows, or notably with 'downward causality' by executed software instructions. We think this is a deeper insight than the one which had been possible for Popper about four decades ago.

Second, extended software comprises more than just digital technical systems. Therefore we propose to call it 'Worldware'. The objects of the quasi-material subworld 1 obtain on the other hand the umbrella term 'Worldstuff'. 'Worldware' implies all intellectual and soul processes, although with very different hardware and architecture. Some of this hardware is even the "porridge" of brains (Alan Turing, 1950). It is important that processes are the main elements and not information. This introduces an operative and constructive view to the non-physical world which is compatible to the technical computer world. It is therefore possible to simulate and reconstruct biological and neurological systems in the computer and to mix digital software and nature software.

As a philosophical result, the subworld 2 extends and the transcending *World 3* shrinks to the intangible (or even hypothetic) objects and processes of some absolute 'Geist' in the sense of Hegel, as are, for example, the creation of arts or love per se.

Third, chance infiltrating the world is a major force which can by no means be neglected (we prefer the neutral German word 'Zufall'). Natural structures generated or influenced by chance are in nature more the rule than the exception.

Lastly, mathematics is here, and in Popper's World, a mysterious and miraculous construction and umbrella. We regard mathematics in some sense both as software and as link to the depths of physics.

The world model and more on the interaction of philosophy and computers (i.e., software) can be found in the book *Wechselwirkung. Wie Prinzipien der Software die Philosophie verändern* (in German), published by Springer Heidelberg in 2016 by the same author.

- Inverse path of development life sciences and computer science
- From spiritual to an operative mode of understanding
- Shannon information does not help in the contrary
- Sorites Problem and gradually growing